\input harvmac
\Title{\vbox{\baselineskip14pt
\hbox{HUTP-96/A036}\hbox{UCB-PTH-96/36}
\hbox{LBNL-39220}\hbox{hep-th/9608079}
}}
{Summing up D-Instantons}
\bigskip
\centerline{Hirosi Ooguri}
\smallskip
\centerline{\it Department of Physics, University of California}
\centerline{\it 366 Le\thinspace Conte Hall, Berkeley, CA 94720, USA}
\centerline{\it and}
\centerline{\it Theoretical Physics Group, Physics Division}
\centerline{\it Ernest Orlando Lawrence Berkeley National
Laboratory}
\centerline{\it University of California, Berkeley,
CA 94720, USA}
\bigskip\centerline{and}\smallskip
\centerline{Cumrun Vafa}
\smallskip
\centerline{\it  Lyman Laboratory of Physics, Harvard
University}
\centerline{\it Cambridge, MA 02138, USA}
\vskip .3in
We investigate quantum corrections to the moduli space for 
 hypermultiplets
for type IIA near a conifold singularity.   We find a unique
quantum deformation based on symmetry arguments which is consistent with
a recent conjecture.  The correction can be interpreted as an infinite
sum coming
from  multiple wrappings of the
Euclidean D-branes  around the vanishing cycle.  
\Date{\it {August, 1996}}

\newsec{Introduction}

Non-perturbative aspects of string theory have been studied
vigorously recently.  These have been both in the
form of solitonic objects as well as instanton corrections
to various physical quantities.  An important class of such objects
is in the form of D-branes or D-instantons
 \ref\poc{J. Polchinski, {\it Combinatorics
of Boundaries in String Theory}, Phys. Rev. {\bf D50} (1994) 6041.},
\ref\pol{J.~Polchinski, {\it Dirichlet Branes and Ramond-Ramond
Charges}, Phys. Rev. Lett. {\bf 75} (1995) 4734.}. 
Most applications to date involve D-branes as solitons.
However one can consider Euclidean p-branes wrapped
around non-trivial cycles of the compactification manifold
to obtain instanton corrections to various physical quantities.
This aspect has been far less studied, however, except
for the considerations of Euclidean membranes
 \ref\bbs{K.~Becker, M.~Becker and A.~Strominger, 
{\it Fivebranes, Membranes and NonPerturbative String Theory},
Nucl. Phys. {\bf B456} (1995) 130.} and Euclidean fivebranes
\ref\witten{E.~Witten, {\it Non-Perturbative Superpotentials
in String Theory}, hep-th/9604030, IASSNS-HEP-96-29.}\ in the
context of M-theory compactifications.
 
In this paper, we study D-brane instanton corrections to the
hypermultiplet
moduli space of type II string compactification on a Calabi-Yau (CY) 
3-fold\foot{Our results also have
a natural interpretation in the context
of M-theory compactifications
near a conifold singularity.}. 
In particular, we examine the moduli space near the conifold
singularity where the non-perturbative aspects
are expected to be crucial.  
In the type IIA  string, the complex moduli of CY 3-fold belong to the
hypermultiplet, and the conifold singularity is realized when
there is a non-trivial 3-cycle in CY whose period 
\eqn\period{ z = \int \Omega }
is small. In the limit $z \rightarrow 0$, the classical hypermultiplet
moduli space develops a singularity, as we will explain in
the next section. 

Before describing the resolution of the conifold singularity in the
hypermultiplet moduli space, let us remind ourselves how a
similar singularity was resolved in the vector multiplet moduli space.
In type IIB string, the complex moduli belong to the vector multiplet and the
same limit $z \rightarrow 0$ generates a singularity in the classical
vector multiplet moduli space. In this case, however, we know that the
moduli space will not be corrected by quantum string effect,
perturbative
or non-perturbative.  In particular the exact leading
singular part of the metric for the
vector multiplet moduli is given by
$$ds^2=-{\rm log}(z\bar z )dzd\bar z .$$
 It was pointed out by Strominger \ref\strominger{
A.~Strominger, {\it Massless Black Holes and Conifolds in String
Theory}, Nucl. Phys. {\bf B451} (1995) 97.}
 that a D3-brane wrapping on the vanishing 3-cycle 
has a mass of order $|z|/\lambda$, where $\lambda$
is the string coupling, and that the conifold singularity
is a reflection of the fact that
 we ignore the light solitonic particle arising from the
D3-brane in string
perturbation theory. If we include it, the low energy effective theory is
 regular even at the conifold point. 

Now let us come back to the hypermultiplet moduli space. Since
the type IIA string does not have a D3-brane, there is no solitonic
state which can become massless at the conifold.  On the other hand,
we may consider the Euclidean
D2-brane  which is wrapped around  the vanishing 3-cycle. 
The hypermultiplet moduli space is not protected against quantum
corrections, and the D2-brane instanton would have an effect of
order ${\rm exp}(-|z|/\lambda)$. It was conjectured in \bbs\ that the
instanton effect should resolve the conifold singularity.
Recently a precise form for this resolution was conjectured
in \ref\BDV{B. Greene, D. Morrison and C. Vafa, {\it A
Geometric Realization of Confinement}, hep-th/9608039.}.
 In this paper
we show that, if we take into account various symmetries,
there is a unique quantum deformation to the conifold singularity in
the classical moduli space and that this result agrees with the
conjecture of \BDV .
 Moreover the modification to the
classical metric is exactly of the form expected for the
multiply wrapped Euclidean  D2-branes (or Euclidean membranes of
M-theory) around the vanishing $S^3$. 
This explicit result, which in effect sums up
the contribution of infinitely many D-instantons, may shed light on
how to sum up D-instantons in other cases as well. 

\newsec{Classical moduli space}

Let us consider the type IIA string on a CY 3-fold $M$. If
$n= {\rm dim} H^{2,1}(M) $, the hypermultiplet moduli space
is complex $(2n+2)$ dimensional; $n$ of which come from
the complex moduli of $M$, $(n+1)$ from the RR 3-form gauge
potential and $1$ from the dilaton and axion $S$. Since we
are interested in the universal behavior of the moduli space
near the conifold limit $z \rightarrow 0$, we will send the
string coupling constant $\lambda \rightarrow 0$
while  keeping $|z|/\lambda$ finite.  In this limit
we may hope to extract a universal deformation of moduli
space, including the important instanton effects of the order
${\rm exp}(-|z|/\lambda)$, which would be
independent of how the vanishing cycle is embedded
in the rest of CY.
 Even though the hypermultiplet moduli
space is quaternionic \ref\cfg{S.~Cecotti, S.~Ferrara and
L.~Girardello, {\it Geometry of Type II Superstrings and
the Moduli of Superconformal Field Theories,} Int. J. Mod. Phys.
{\bf A4} (1989) 2475.}\ref\fs{S.~Ferrara and S.~Sabharwal,
{\it Quaternionic Manifolds for Type II Superstring Vacua
of Calabi-Yau Spaces,} Nucl. Phys. {\bf B332} (1990) 317.},
in the limit we are considering the relevant piece of the
singularity is a hyperk\"ahler manifold of real dimension 4.
The complex moduli $z$ is paired with 2 real coordinates
$x$ and $t$ which are expectation values of the RR 3-form
corresponding to the vanishing cycle  and its dual respectively. 
In the following, we will concentrate on the subspace of moduli space
spanned by $z$ and $(t,x)$.

Since the RR charges carried by D-branes are quantized, 
the moduli space must be periodic in the RR 3-forms $x$ and
$t$.  We normalize them so that each has period 1.
 Moreover there is a monodromy action on $H_3(M)$ as
$z$ goes around the conifold point, and this mixes 
$(t,x) \rightarrow (t+x,x)$. Thus the moduli space geometry 
near the conifold is described by the elliptic fibration
\eqn\ell{ \tau(z) = {1 \over 2 \pi i} {\rm log} z .}
This is similar to the situation of the stringy cosmic string
\ref\gsvy{B.R.~Greene, A.~Shapere, C.~Vafa and S.-T.~Yau,
{\it Stringy Cosmic Strings,} Nucl. Phys. {\bf B337} (1990) 1.}.
In fact at weak coupling,
 the leading singularity in the classical moduli space metric
computed using the result of \fs\ agrees with that of \gsvy.
The K\"ahler form for the classical metric is given by
\eqn\class{  k = 
     \partial \bar{\partial} \left( 
     {(\zeta - \bar{\zeta})^2 \over 2 (S+\bar{S})\tau_2 } 
   \right) +\tau_2 dz d\bar{z} }
where $\zeta = t + \tau x$ and $\tau_2 = {\rm Im} \tau(z)$.
The metric has a $U(1)_t \times U(1)_x$ translational 
invariance in $t$ and $x$.
This is to be expected since
there is no perturbative string state which carries the RR charges.  
Since $(S + \bar{S})$ is the dilaton from the NS-NS
sector, $(S + \bar{S}) \sim 1/\lambda^2$ where $\lambda$
is the string coupling constant (for
a precise definition of the string coupling
constant in the present context see \ref\bs{N.
Berkovits and W. Siegel, {\it Superspace Effective
Actions for 4D Compactifications of Heterotic and Type II
Superstrings}, Nucl. Phys. {\bf B462} (1996) 213.}). In the following,
we will use $1/\lambda^2$ in place of $(S+\bar{S})$.

The metric of the classical moduli space discussed in the above
is singular at the conifold point $z=0$ as shown
in \gsvy . In the neighborhood of 
$z=0$, however, we expect large instanton effects due to 
Euclidean
D2-branes wrapping the vanishing 3-cycle.  It has been conjectured
in \bbs\ that such effects would resolve the singularity at $z=0$. 
In \BDV\ based on some field theory considerations, it was conjectured
more precisely that the exact corrected metric is the unique
hyperk\"ahler metric where the K\"ahler class of the elliptic
fiber is $\lambda ^2$. In the following, we will {\it derive}
the form of the corrected metric based on symmetry
considerations for Euclidean membranes wrapped around vanishing
cycle and the assumption of resolution of the singularity \bbs\
and find agreement with the conjecture in \BDV .  This
also leads us to an explicit realization
of the metric for which the classical part and quantum
corrections can be identified.  Moreover the quantum
corrections can be naturally reinterpreted
as D-instanton contributions to the metric.

\newsec{Quantum moduli space}

In order to exhibit the symmetries of the metric \class, it is
convenient to rewrite it as
$$ ds^2 = {\lambda^2 \over \tau_2}
     (dt + \tau(z) dx)(dt + \bar{\tau}(\bar{z}) dx)  + \tau_2 dz d\bar{z}
     . $$
Note that the metric has $U(1)_t \times U(1)_x$
symmetries corresponding to the translations in $t$ and $x$,
as we explained in section 2. 
One then recognizes that it takes
the form of the ansatz \ref\hawking{
S.W.~Hawking, {\it Gravitational Instantons},
Phys. Lett. {\bf 60A} (1977) 81.} \ref\eh{
T.~Eguchi and A.J.~Hanson, {\it Selfdual Solutions to Euclidean
Gravity,} Ann. Phys. {\bf 120} (1979) 82.}
for a self-dual metric:
\eqn\selfdual{ ds^2 = \lambda^2
 \left[ V^{-1} (dt - {\bf A}\cdot {\bf dy}) ^2+ V d{\bf y}^2 \right] , }
with ${\bf y} = (x, z/\lambda, \bar{z}/\lambda)$
and
$$\eqalign{
 V &= \tau_2 = {1 \over 4 \pi} {\rm log} \left( 
            {1 \over z \bar{z}} \right)  \cr 
  A_x & = -\tau_1 = {i \over 4 \pi } {\rm log} \left(
             {z \over \bar{z}} \right),~~
  A_z = 0, ~~ A_{\bar{z}} = 0 \cr}$$ 
This metric is singular at $z=0$. Moreover we are taking
$t$ and $x$ to be periodic with period $1$. 

Now let us discuss how the quantum
corrections could modify the metric, paying attention
to the fate of the $U(1)_t\times U(1)_x$ translational
invariance. 
Since the variable $x$ 
corresponds to the expectation value
of RR 3-form on the vanishing cycle, 
the D2-instanton wrapping on 
it would break the translational invariance in the $x$-direction.
In particular if we consider a Euclidean 2-brane wrapped $m$
times around
$S^3$ it couples to RR expectation value on it and gives
us a factor of ${\rm exp}(2\pi i mx)$.  Note that this
is still consistent with periodicity of $x$, i.e.
the $U(1)_x$ has been broken to ${\bf Z}$.
On the other hand 
$t$ couples to a cycle dual to the vanishing $S^3$, and its translational
invariance will not be broken, as we are considering  a limit where
the dual period is not vanishing\foot{Note that if we had been
considering a case where the dual cycle also has vanishing
period the translation in $t$ would also be broken.  This should
be interesting to study.}
 and thus is irrelevant in
the leading order as $\lambda \rightarrow 0$.
Thus it is appropriate to work in the ansatz
\selfdual.

There are various requirements that the potential $V$ has to
satisfy:

\hskip -.3in
(1) The metric is hyperk\"ahler if and only if $V$ and ${\bf A}$
obey
$$ V^{-1} \Delta V = 0,~~
   \nabla V = \nabla \times {\bf A}.,$$
where
$$  \Delta = \partial_x^2 + 4 \lambda^2 
      \partial_z \bar{\partial}_{\bar{z}} .$$
The factor $V^{-1}$ in the first equation means that
we allow delta-function singularities in $\Delta V$. 
Thus we can think of $V$ as the electro-magnetic scalar
potential for a collection of charges in 3-dimensions. 

\hskip -.3in
(2) For large $z$ when the instanton effects are suppressed, 
the metric should reduce to the classical one. 
$$  V \rightarrow {1 \over 4 \pi} {\rm log} \left( 
 {1 \over z \bar{z}} \right) ~~~~( |z| \rightarrow \infty). $$

\hskip -.3in
(3) The metric should be periodic,
but not translationally invariant, in $x$ with the period $1$.

\hskip -.3in
(4) Since the Calabi-Yau geometry near the conifold is 
invariant under the phase rotation of $z$ and that 
Euclidean membranes only probe the overall size $|z|$
of $S^3$,
 the $dz d \bar{z}$
part of the moduli space metric should be independent of
the phase.  This means that the potential $V$ is a function
of $x$ and $|z|$ only.

\hskip -.3in
(5)  For a single conifold we assume the quantum
metric has no singularity.  This means in particular
that the singularities of $V$ must be such that they can 
be removed by appropriate coordinate transformation.

The conditions (1), (3) and (4)  means that we are to find the 
electro-magnetic potential $V$ which is periodic in $x$ and axial
symmetric in the $z$-plane. The condition (2) says that
the electric charges are distributed near the axis $z=0$ and its density
per unit length in $x$ is 1. The condition (5) requires that
these charges to be quantized in the unit of 1
and that no two charges are at the same point.  In particular
if we have $N$ charges at the same point
the space
will develop ${\bf C}^2/{\bf Z}_N$ singularity.
  There is a unique solutions satisfying
these conditions, and it is given by
\eqn\taubnut{  V ={1 \over 4 \pi} \sum_{n= -\infty}^\infty \left(
    {1 \over  \sqrt{ (x - n)^2 +  z \bar{z}/ \lambda^2 }  }
  - {1 \over |n|} \right) + {\rm const}.}

To exhibit the D-instanton effects, it is convenient to take
the Poisson resummation of this potential. We then find
\eqn\fourier{ V = {1 \over 4 \pi} {\rm log}\left({ \mu^2 \over
        z \bar{z}} \right) +
  \sum_{m\neq 0} {1 \over 2\pi} e^{2 \pi i m x} K_0\left(2 \pi 
     {| m z| \over  \lambda}  \right) , }
where $\mu$ is some constant. 
$K_0$ is the modified Bessel function, whose appearance
is natural in the axially symmetric potential problem. 
By construction, the metric is regular at $z=0$ and reduces
to the classical one $V \sim {1 \over 4 \pi} {\rm log} (1/z\bar{z})$
for $|z| \rightarrow \infty$.

\newsec{Interpretation}

When $z$ is large, we can use the asymptotic formula of
the Bessel function to expand \fourier\ as
\eqn\largez{ 
\eqalign{ V =   {1 \over 4 \pi } {\rm log}\left({\mu^2 \over z \bar{z}}
\right)
+ \sum_{m \neq 0}&  {\rm exp}\left[ - 2 \pi \left( 
   {|mz| \over \lambda} - imx \right)\right]
\times \cr &\times 
 \sum_{n=0}^\infty{ \Gamma({1 \over 2}+n) \over
2 \sqrt{\pi} n!
   \Gamma({1 \over 2} -n)} 
  \left({ \lambda \over 4 \pi|mz|}\right)^{n+{1 \over
   2}}. \cr}}
Notice that the correction to the classical term
${1 \over 4 \pi}{\rm log}(1/z\bar{z})$ is
exponentially suppressed by the factor
${\rm exp}[- 2 \pi ( |mz| / \lambda - i mx)] $.
This is exactly what we expect for the instanton effect
due to D2-branes wrapping the vanishing $S^3$. 
The D2-instanton configuration should
preserve $1/2$ of the spacetime supersymmetry, which means 
in particular that the volume form on the membrane worldvolume 
is proportional to the holomorphic 3-form $\Omega$ \bbs.
Thus $2 \pi |mz|/\lambda$ in the exponent is nothing but
the Born-Infeld action for the $m$-instanton
($m$ times wrapping of $S^3$).
 Since $x$ is the integral of the
RR 3-form on $S^3$, the second term $2 \pi i m x$
in the exponent describes the coupling of the
D2-brane to the RR field. 

Note that this result implies a number of things:
First of all there is no perturbative correction
to the leading singularity of hypermultiplet moduli
near the conifold singularity.  Secondly, perhaps surprisingly,
{\it all} instanton numbers are present for the correction
to the metric.  This is in contract with the count
of stable solitons in the type IIB near the conifold where
the multiply wrapped state is not expected to be
stable \strominger \ref\bsv{M. Bershadsky, V. Sadov
and C. Vafa, {\it D-branes and Topological Field
Theories}, Nucl. Phys. {\bf B463} (1996) 420.}.
Thirdly for each D-instanton we have an infinite
`perturbative' sum.  It would be interesting
to connect this to perturbative string computations
around the D-instanton background.
In this connection the open topological string
theory on $T^* S^3$ \ref\witc{E. Witten, {\it Chern-Simons Gauge
Theory as a String Theory}, hep-th/9207094.}\
may be relevant\foot{This suggestion arose during
conversations with C. Imbimbo and K.S. Narain.} 
(See in particular \ref\ol{M.~O'Loughlin, {\it
Chern-Simons from Dirichlet 2-Brane Instantons},
hep-th/9601179, to appear in Phys. Lett. B.}).
  It is also surprising
that the power of the coupling $\lambda$ is shifted
from an integer by $1/2$.  This may be related to a
precise definition of the coupling constant \bs .

Note that if we consider the case of $N$ vanishing
3-cycles instead of 1, then our considerations
naturally lead to $V\rightarrow NV$.  This space
will have ${\bf C}^2/{\bf Z}_N$ singularity.
 This is in 
agreement with the conjecture in \BDV\ and the fact
that the Euclidean membranes treat each vanishing $S^3$
independently of each other.

\bigskip

\bigskip

\centerline{\bf Acknowledgments}

We would like to thank T. Eguchi,
 B. Greene, K. Intriligator,
D. Morrison, J. Polchinski and A. Strominger for
valuable discussions.
We gratefully acknowledge the hospitality of the Aspen Center for
Physics, where this work has been done.
H.O. is supported in part by NSF grant PHY-951497
and DOE grant DE-AC03-76SF00098.
C.V. is supported in part by NSF grant PHY-92-18167.

\listrefs

\end